\documentclass[aps,prl,twocolumn,showpacs,superscriptaddress,groupedaddress]{revtex4}
\usepackage{mathbbol}
\usepackage{amsmath}
\usepackage{amsfonts}
\usepackage{amssymb}
\usepackage{graphicx}
\usepackage{color}
\usepackage{morefloats}
\usepackage{multirow}
\usepackage{relsize}
\usepackage{amsthm}
\usepackage{ytableau}
\usepackage{tikz}
\usepackage{mathdots}
\usetikzlibrary{decorations.pathreplacing}

\begin{document}

\newcommand{\ket}[1]{\left |#1 \right \rangle}
\newcommand{\bra}[1]{\left  \langle #1 \right |}
\newcommand{\braket}[2]{\left \langle #1 \middle | #2 \right \rangle}
\newcommand{\ketbra}[2]{\left | #1 \middle \rangle \middle \langle #2 \right |}

\newcommand{\tikznode}[3][inner sep=0pt]{\tikz[remember
picture,baseline=(#2.base)]{\node(#2)[#1]{$#3$};}}

\newtheorem{conj}{Conjecture}
\newtheorem{thm}{Theorem}

\widetext

\title{Mean Field Approximation for Identical Bosons on the Complete Graph}
\author{Alexander Meill and David A. Meyer}
\date{\today}

\begin{abstract}
Non-linear dynamics in the quantum random walk setting have been shown to enable conditional speedup of Grover's algorithm.  We examine the mean field approximation required for the use of the Gross-Pitaevskii equation on identical bosons evolving on the complete graph.  We show that the states of such systems are parameterized by the basis of Young diagrams and determine their one- and two party marginals.  We find that isolated particles are required for good agreement with the mean field approximation, proving that without isolated particles the matrix fidelity agreement is bounded from above by $1/2$.
\end{abstract}

\maketitle

\section{Introduction}

It has been shown that certain computational tasks, such as search \cite{Grover:1996rk} and factorization \cite{Shor:1994:AQC:1398518.1399018}, can be performed faster by a quantum computer than a classical one.  This speed up over the classical information scheme can be attributed to hallmark properties of quantum mechanics, including superposition and entanglement.  It is concievable, then, that continuing to add or tweak the features of the computation model might bring further improvements.  The $\mathcal O( \sqrt N)$ scaling of Grover's search algorithm, for example, has been found to be optimal in the quantum model \cite{doi:10.1137/S0097539796300933}, whose dynamics are linear, as described by the Schrodinger Equation.  If we instead assume non-linear dynamics, however, randomly walking particles can conditionally search with even better scalings \cite{PhysRevA.89.012312}, depending on the non-linearity, even approaching constant time scaling for certain models.

To make practical use of the speedups provided these approximations to the quantum model, we need to confirm their physical significance.  The Gross-Pitaevskii Equation (GPE) \cite{Gross1961}, for instance, is a non-linear approximation which is particularly appropriate for describing the dynamics of Bose-Einstein condensates \cite{Rogel_Salazar_2013}.  If we accept and apply the GPE to quantum random walks on the complete graph, search has been shown to conditionally scale like $\mathcal O(N^{1/4})$ \cite{Meyer_2013}.

In order for the GPE to be a reasonal approximatin to the Schrodinger Equation, though, a number of assumptions must be satisfied.  Assuming many particles and short-ranged interactions is relatively benign, and enforcing permutation symmetry is simply a matter of restricting to identical Bosons.  The more challenging assumptions to motivate physically are the mean field approximation,
\begin{equation}
\rho_2 \left(t,x_1,x_2;x_1'x_2' \right) \approx \rho_1 \left(t,x_1;x_1' \right) \otimes \rho_1 \left(t,x_2;x_2' \right),
\end{equation}
and approximate purity of the single-party reduced state,
\begin{equation}
\rho_1 \left(t,x;x' \right) \approx \psi \left(t,x \right) \psi^* \left( t,x' \right),
\end{equation}
where $\rho_2$ and $\rho_1$ are the continuous space one- and two-party reduced states respectively, noting that the permutation symmetry of the overall state makes the choice of specific parties labels irrelevant.  These are strong restrictions to make to the overall Hilbert space, but are properties natural to Bose-Einstein condensates \cite{Rogel_Salazar_2013}, which make them a common physical medium for the continuous-space GPE.

To extend this analysis to quantum random walks, the GPE, along with its requisite assumptions, must be translated to a discrete space description.  In particular, the mean field approximation and pure single-party marginal assumptions become
\begin{align}\label{MFA}
\rho_2 &\approx \rho_1 \otimes \rho_1 \\
\rho_1 &\approx \ketbra{\psi}{\psi},
\end{align}
where the states are now described by finite-dimensional vectors and density matrices.  The aim of this paper is to examine the validity of these assumptions, particularly the mean field approximation (\ref{MFA}), for quantum random walks of permutation-invariant particles on the complete graph.  Some entanglement properties of such states have been examined previously \cite{RobinReuvers:2019xew}, but we provide a more targeted approach.

At first glance it seems as if the symmetries associated to the permutation-invariance and the complete graph may be enough to guarantee (\ref{MFA}) because it is, in some sense, a measure of entanglement.  Certainly if $\rho_2$ is entangled then (\ref{MFA}) will not be satisfied, and potentially the converse is true; the less entangled the state is, the greater the agreement with (\ref{MFA}).  The monogamy of entanglement then suggests that the symmetric sharing of pairwise entanglement in the overall state will decay with the number of particles, $n$.  Permutation-invariance alone in multi-qubit states bounds the pairwise Concurrence \cite{PhysRevLett.80.2245} by $\mathcal C_{i,j} \leq 2/n$ \cite{PhysRevA.62.050302}.

In this paper we begin by fully expressing the symmetries of permutation-invariance and the complete graph, then finding the resulting Young diagram basis for the symmetrized states.  We then measure the validity of the mean field approximation on those states through the matrix Fidelity.  We perform exact fidelity calculations on a subset of  single basis elements, then bound the fidelity for a larger subset of states.

\section{Party-Site Symmetry}

Consider $n$ particles evolving on a complete graph of $d$ sites.  To describe the state of such a system, we can index the site position of particle $j$ by $i_j$, and therefore the overall state can be described by $n$ qu$d$its, or, a vector in $\mathbb C_d^{\otimes n}$,
\begin{equation}\label{ogPSS}
\ket{\psi} = \sum_{i_1 = 1}^{d} \ldots \sum_{i_n=1}^{d} a_{i_1 \ldots i_n} \ket{i_1 \ldots i_n}.
\end{equation}
To this state we then want to enforce constraints which reflect the symmetries of both:
\begin{itemize}
\item \emph{Party:}  Because the particles in question are identical, we want to enforce that any permutation of their labels leaves the state unchanged.
\item \emph{Site:}  Because the particles are walking on a \emph{complete} graph, the sites themselves of that graph are identical, and any permutation of their labels would leave the graph, and therefore the state, unchanged.
\end{itemize}
We can formalize these combined symmetries by defining that a state,
\begin{equation}
\ket{\psi} = \sum_{i_1 = 1}^{d} \ldots \sum_{i_n=1}^{d} a_{i_1 \ldots i_n} \ket{i_1 \ldots i_n},
\end{equation}
is party-site symmetric (PSS) if
\begin{eqnarray}\label{Usym}
U_{\mu} \ket{\psi} &=& \ket{\psi} \quad \forall \quad \mu \in S_n \\ \label{Vsym}
V_{\nu}^{\otimes n} \ket{\psi} &=& \ket{\psi} \quad \forall \quad \nu \in S_d,
\end{eqnarray}
where $U_{\mu}$ is the unitary representation of $\mu$, which permutes the party labels,
\begin{equation}
U_{\mu} \ket{i_1 \ldots i_N} = \ket{ \mu \left( i_1 \ldots i_N \right)},
\end{equation}
while $V_{\nu}$ is the unitary representation of $\nu$, which permutes the basis (site) labels,
\begin{equation}
V_{\nu} \ket i = \ket{\nu(i)}.
\end{equation}
The individual symmetries associated to party (\ref{Usym}) and site (\ref{Vsym}) will be referred to as $U$ and $V$ symmetries respectively.  For PSS states we expect that many of the coefficients, $a_{i_1 \ldots i_n}$, are constrained to be equal by the $U$ and $V$ symmetries, leaving some much smaller basis for the subspace.  The $U$ symmetry implies that the ordering of $i_1$-$i_n$ does not matter.  The $V$ symmetry then implies that the collective index values themselves can be freely permuted. Given these constraints, the only actually distinguishing feature of a given $a_{i_1 \ldots i_n}$, and the elements it is grouped with, is the partitioning of shared indices.  For example, $a_{1,2,2,3,4,2,3}=a_{2,2,2,3,3,1,4}=a_{i_1,i_1,i_1,i_1,i_2,i_2,i_3,i_4}$ would be grouped under the label, $a_{3,2,1,1}$, where now the subscripts denote the number of parties who share a given index.  One can recognize that such a grouping and labeling can be expressed in a young diagram,
\begin{widetext}
\begin{center}
\ytableausetup{centertableaux}
\begin{ytableau}
\; & & & \\
\; & \\
\\
\end{ytableau}
$=\left \{ \mu^{\vphantom{U^U}} \left( \nu(1^{\vphantom{U^U}}),\nu(2),\nu(2),\nu(3),\nu(4),\nu(2),\nu(3) \right)^{\vphantom{U^U}} \; \; \middle  | \; \; \mu \in S_n, \; \nu \in S_d \right \}$.
\end{center}
\end{widetext}
In general, the number of rows in a Young diagram indicates that each of the elements in the set has that many distinct indices.  The number of blocks in a row indicates how many parties share that index.  Naturally, the total number of blocks is $n$, and there can be at most $d$ rows in a Young diagram.  With the interpretation of Young diagrams established, we then find that they serve as an orthonormal basis for pure PSS states,
\begin{equation}\label{PSS}
\ket{\psi} = \sum_{y \in \mathcal Y(n,d)} a_y \ket y,
\end{equation}
where $\mathcal Y(n,d)$ is the set of Young diagrams with $n$ blocks and at most $d$ rows, and $\ket y$ is a normalized equal superposition of computational basis elements belonging to the set described by the Young diagram, $y$.

In evaluating the validity of the mean field approximation for a PSS state we will have to perform the partial trace on (\ref{PSS}) to find $\rho_1$ and $\rho_2$.  Thankfully, the $U$ and $V$ symmetries greatly simplify that process.  Consider the reduced state, $\rho_k$, obtained by tracing out the last $n-k$ parties,
\begin{widetext}
\begin{align}
\rho_k =& \text{Tr}_{\bar k} \left( \ketbra{\psi}{\psi} \right) \\
 =& \sum_{l_{k+1} \ldots l_{N}} \; \sum_{i_{1} \ldots i_k} \; \sum_{j_{1} \ldots j_k} a_{i_1 \ldots i_k l_{k+1} \ldots l_N} \ketbra{i_1 \ldots i_k}{j_1 \ldots j_k} a_{j_1 \ldots j_k l_{k+1} \ldots l_N}^*.
\end{align}
\end{widetext}
Now consider $V_{\nu^{-1}}$ for some $\nu \in S_d$, acting on $\rho_k$,
\begin{widetext}
\begin{eqnarray}
V_{\nu^{-1}} \rho_k V_{\nu^{-1}}^{\dagger} &=& \sum_{i, j, l} a_{\nu \left(i_1\right) \ldots \nu \left(i_k\right) l_{k+1} \ldots l_N} \ketbra{i_1 \ldots i_k}{j_1 \ldots j_k} a_{\nu \left(j_1\right) \ldots \nu \left(j_k\right) l_{k+1} \ldots l_N}^* \quad \quad \quad \\
&=& \sum_{i, j, l} a_{\nu \left(i_1\right) \ldots \nu \left(i_k\right) \nu \left(l_{k+1}\right) \ldots \nu \left(l_N\right)} \ketbra{i_1 \ldots i_k}{j_1 \ldots j_k} a_{\nu \left(j_1\right) \ldots \nu \left(j_k\right) \nu \left(l_{k+1}\right) \ldots \nu \left(l_N\right)}^* \quad \quad \quad \\
&=& \sum_{i, j, l}  a_{i_1 \ldots i_k l_{k+1} \ldots l_N} \ketbra{i_1 \ldots i_k}{j_1 \ldots j_k} a_{j_1 \ldots j_k l_{k+1} \ldots l_N}^* \\
&=& \rho_k.
\end{eqnarray}
\end{widetext}
Likewise consider $U_{\mu^{-1}}$ for $\mu \in S_k$.  We can also extend $\mu \otimes \mathbb 1_{n-k} \in S_n$ as the permutation which acts on the first $k$ parties by $\mu$ and leaves the traced over parties fixed.  Now examine
\begin{widetext}
\begin{eqnarray}
U_{\mu^{-1}} \rho_k U_{\mu^{-1}}^{\dagger} &=& \sum_{i, j, l} a_{\mu \left(i_1 \ldots i_k\right) l_{k+1} \ldots l_N} \ketbra{i_1 \ldots i_k}{j_1 \ldots j_k} a_{\mu \left(j_1 \ldots j_k\right) l_{k+1} \ldots l_N}^* \quad \quad \quad \\
&=& \sum_{i, j, l} a_{\mu \otimes \mathbb 1_{N-k} \left(i_1 \ldots i_k l_{k+1} \ldots l_N \right)} \ketbra{i_1 \ldots i_k}{j_1 \ldots j_k} a_{\mu \otimes \mathbb 1_{N_k} \left(j_1 \ldots j_k l_{k+1} \ldots l_N \right)}^* \quad \quad \quad \\
&=& \sum_{i, j, l}  a_{i_1 \ldots i_k l_{k+1} \ldots l_N} \ketbra{i_1 \ldots i_k}{j_1 \ldots j_k} a_{j_1 \ldots j_k l_{k+1} \ldots l_N}^* \\
&=& \rho_k.
\end{eqnarray}
\end{widetext}
Even more interesting is that acting on only one side by $U$ would likewise leave $\rho_k$ invariant because $\mu$ can be freely extended to $\mu \otimes \mathbb 1_{N-k}$ for either the bra or the ket individually.  This is not true for the $V$ symmetry, where absorbing $\nu$ into the sum in $l$ has to affect both the bra and the ket simultaneously.  Altogether then we have
\begin{eqnarray}
U \rho_k = \rho_k U = \rho_k \\
V \rho_k V^{\dagger} = \rho_k.
\end{eqnarray}

These symmetries allow us to greatly constrain the elements of $\rho_1$ and $\rho_2$, leaving us with a fairly simple parametrization of the two matrices.  Starting with $\rho_1$, the $V$ symmetry implies that $\left \{ \rho_1 \right \}_{i,j} = \left \{ \rho_1 \right \}_{\nu(i), \nu(j)}$ for any $\nu \in S_d$.  This then equates all the diagonal elements as $\left \{ \rho_1 \right \}_{i,i} = \frac 1d$ and the off diagonal elements as $\left \{ \rho_1 \right \}_{i,j} = A$ for all $i \neq j$.  Since the $V$ symmetry equates $\left \{ \rho_1 \right \}_{i,j}= \left \{ \rho_1 \right \}_{j,i}$, the hermiticity of $\rho_1$ then implies that $A \in  \mathbb R$.  The same application of the $U$ and $V$ symmetries along with hermiticity constrain the following elements of $\rho_2$, in which it is implied that $i$, $j$, $k$, and $l$ are distinct,
\begin{align}
B_1 =& \left \{ \rho_2 \right \}_{i j, k l}  \\
B_2 =&\left \{ \rho_2 \right \}_{i i, k l}  \\
B_3 =&\left \{ \rho_2 \right \}_{i j, i l} = \left \{ \rho_2 \right \}_{i j, l i} = \left \{ \rho_2 \right \}_{j i, i l} = \left \{ \rho_2 \right \}_{j i, l i}  \\
B_4 =&\left \{ \rho_2 \right \}_{i i, k k} \\
B_5 =&\left \{ \rho_2 \right \}_{i i, i l} = \left \{ \rho_2 \right \}_{i i, l i}  \\
C_1 =&\left \{ \rho_2 \right \}_{i j, i j} = \left \{ \rho_2 \right \}_{i j, j i} \\
C_2 =&\left \{ \rho_2 \right \}_{i i, i i}
\end{align}
where $B_1$, $B_3$, $B_4$, $C_1$, and $C_2$ are real, while $B_2$ and $B_5$ are complex.  Notably, we can relate $A$ to the parameters of $\rho_2$ by
\begin{align}
A &= \left \{ \rho_1 \right \}_{i,j} \\
&= \left \{ \text{Tr}_2 \left ( \rho_2 \right ) \right \}_{i,j}\\
&= \sum_{k=1}^d \left \{ \rho_2 \right \}_{ik,jk} \\
&= (d-2)B_3 + B_5 + B_5^* \\
&= (d-2)B_3 + 2 \Re \left (B_5 \right ).
\end{align}
We can also use the normalization of $\rho_2$ to find $dC_1+d(d-1)C_2=1$.

\section{Exact Fidelity Calculations}

Having examined $\rho_1$ and $\rho_2$ for PSS states, we must now choose a metric by which to measure the agreement with (\ref{MFA}).  We have chosen to use the matrix Fidelity \cite{doi:10.1080/09500349414552171},
\begin{equation}
F(A,B) = \left[ \text{Tr} \sqrt{\sqrt A B \sqrt A} \right]^2.
\end{equation}
The fidelity is a common choice in quantum information theory as a generalization of the pure state inner product to mixed states.  Applied to the mean field approximation, let us label
\begin{equation}
F(\ket \psi) = \left [ \text{Tr} \sqrt M \right]^2,
\end{equation}
where $M = \sqrt{\rho_1 \otimes \rho_1} \rho_2 \sqrt{\rho_1 \otimes \rho_1}$.  Unfortunately, for the most general PSS state (\ref{PSS}), the resulting $M$ matrix is analytically challenging to diagonalize or find the trace of its square root.  In some simple cases, however, the fidelity can be determined exactly.  Consider the following set of PSS states described by single, rectangular Young diagram basis elements,
\ytableausetup{centertableaux,boxsize=6mm}
\[y(k) = \quad \quad
\begin{ytableau}
\tikznode{a3}{~}\; & & \none[\dots] & &  \\
\; & \none & \none & \none & \\
\none[\vdots] & \none & \none & \none & \none[\vdots]\\
\; & \none & \none & \none & \\
\tikznode{a1}{~} & & \none[\dots] & & \tikznode{a2}{~}
\end{ytableau}.
\]
\tikz[overlay,remember picture]{%
\draw[decorate,decoration={brace},thick] ([yshift=-3.5mm,xshift=3mm]a2.south east) -- 
([yshift=-3.5mm,xshift=-3mm]a1.south west) node[midway,below]{$\frac nk$};
\draw[decorate,decoration={brace},thick] ([yshift=-2.5mm,xshift=-4mm]a1.south west) -- 
([yshift=4.5mm,xshift=-3.2mm]a3.north west) node[midway,left]{$k$};
}
\vspace{5mm}

\noindent Expressed as a state in the computational basis,
\begin{equation}
\ket{y(k)} = \mathcal A_k^{- \frac 12} \sum_{\mu \in S_n} \; \sum_{i_1 < \ldots < i_k} U_{\mu} \bigotimes_{j=1}^k \ket{i_j}^{\otimes \frac nk},
\end{equation}
where $\mathcal A_k$ is a normalization constant equal to the number of computational basis elements present in $\ket{y(k)}$.  It evaluates to
\begin{equation}
\mathcal A_k = \binom{d}{k} \frac{n!}{{\left[ \left(\frac nk \right) !\right]}^k} = \left(\frac{k!(d-k)! \left[ \left( \frac nk \right)! \right]^k}{d! \, n!} \right)^{-1}.
\end{equation}
The first step in calculating $F( \ket{y(k)})$ is the determination of the components of $\rho_2$ and $\rho_1$.  Consider first
\begin{align}
\rho_1 &= \mathcal A_k^{-1} \sum_{i, k_2 \ldots k_n \in y} \sum_{j, k_2 \ldots k_n \in y} \ketbra ij \\
&=\mathcal A_k^{-1} \sum_{i,j} \mathcal N^{(y)}_{i,j} \ketbra ij,
\end{align}
where $\mathcal N_{i,j}^{(y)}$ is the number of strings, $(k_2 \ldots k_n)$, for which both $(i \, k_2 \ldots k_n)$ and $(j \, k_2 \ldots k_n)$ are contained in the set associated to $y$.  We can analogously define $\mathcal N_{ij,kl}^{(y)}$ such that
\begin{equation}
\rho_2 = \mathcal A_k^{-1} \sum_{i,j,k,l} \mathcal N_{ij,kl}^{(y)} \ketbra{ij}{kl}.
\end{equation}
Determining each of the $\mathcal N$ for the family of $y(k)$ is a simple counting/combinatorics exercise.  The results are the following, where it is assumed that $i$, $j$, $k$, and $l$ are distinct,
\begin{align}
\mathcal N_{i,j}^{(y(k))} &= \delta(k-n) \frac{(d-2)!}{(d-n-1)!} \\
\mathcal N_{ii,ii}^{(y(k))} &= \begin{cases} 0 & k=n \\ \binom{d-1}{k-1} \frac{(n-2)!}{ \left( \frac nk-2 \right)! \left[ \left( \frac nk \right)! \right ]^{k-1}} & k\neq n \end{cases} \\
\mathcal N_{ij,ij}^{(y(k))} &= \binom{d-2}{k-2} \frac{(n-2)!}{\left[ \left( \frac nk-1 \right)! \right]^2 \left[ \left( \frac nk \right)! \right ]^{k-2}} \\
\mathcal N_{ii,jj}^{(y(k))} &= \delta \left(k-\frac n2 \right) \binom{d-2}{\frac n2-1} \frac{(n-2)!}{ 2^{\frac n2-1}} \\
\mathcal N_{ij,ik}^{(y(k))} &= \delta(k-n) \frac{(d-3)!}{(d-n-1)!} \\
\mathcal N_{ij,kl}^{(y(k))} &= \delta(k-n) \frac{(d-4)!}{(d-n-2)!},
\end{align}
while $\mathcal N_{ii,ij}^{(y(k))}=\mathcal N_{ii,jk}^{(y(k))}=0$.  Dividing by $\mathcal A_k$ then finally gives the components of each reduced density matrix,
\begin{align}
A &= \delta(k-n) \frac{d-n}{d(d-1)} \\
\left \{ \rho_2 \right \}_{ii,ii} &= \frac{n-k}{d \, k(n-1)} \\
\left \{ \rho_2 \right \}_{ij,ij} &= \frac{n(k-1)}{d (d-1) k(n-1)} \\
B_4 &= \delta \left(k- \frac n2 \right) \frac{d- \frac n2}{d(d-1)(n-1)} \\
B_3 &= \delta(k-n) \frac{d-n}{d(d-1)(d-2)} \\
B_1 &= \delta(k-n) \frac{(d-n)(d-n-1)}{d(d-1)(d-2)(d-3)} \\
B_2 &=B_5=0
\end{align}
From here there are three major cases to consider:  $k<n/2$, $k=n/2$, and $k=n$.  Starting with the $k<n/2$ case we have $\rho_1= d^{-1} \mathbb 1_d$ and
\begin{align}
\rho_2 =& \frac{1}{d \, k (n-1)} \biggr[ (n-k) \sum_i \ketbra{ii}{ii} \\ \notag &+ \frac{n(k-1)}{d-1} \sum_{i \neq j} \ketbra{ij}{ij} + \ketbra{ij}{ji} \biggr],
\end{align}
which leads to
\begin{align}
M =& \frac{1}{d^3 k (n-1)} \biggr[ (n-k) \sum_i \ketbra{ii}{ii} \\ \notag &+ \frac{n(k-1)}{d-1} \sum_{i \neq j} \ketbra{ij}{ij} + \ketbra{ij}{ji} \biggr].
\end{align}
Given that we will be tracing this matrix after finding its square root, we can jointly reorder the rows and columns together.  Doing so yields the convenient representation,
\begin{equation}
M = \frac{n(k-1)}{d^4k(n-1)} \left[  \left( \begin{array}{c c} 1 & 1 \\ 1 & 1 \end{array} \right)^{\oplus d(d-1)/2} \oplus \frac{d(n-k)}{n(k-1)} \, \mathbb 1_d \right],
\end{equation}
which we be easily diagonalized and square rooted,
\begin{align}
\sqrt M =& \sqrt{\frac{n(k-1)}{d^4k(n-1)}} \biggr[ \left( \begin{array}{c c} \sqrt 2 & 0 \\ 0 & 0 \end{array} \right)^{\oplus d(d-1)/2} \\ \notag &\oplus \sqrt{\frac{d(n-k)}{n(k-1)}} \, \mathbb 1_d \biggr].
\end{align}
And finally we can find $F( \ket{y(k<n/2)})$,
\begin{align}
F = \frac{ \left( (d-1) \sqrt{n(k-1)}+ \sqrt{2d(n-k)} \right)^2}{2d^2k(n-1)}.
\end{align}
As $d \to \infty$, this simplifies to
\begin{equation}
F( \ket{y(k<n/2)}) = \frac{n(k-1)}{2k(n-1)}.
\end{equation}

Moving to the $k=n/2$ case, the same analysis arrives at
\begin{equation}
F = \frac{\left( \sqrt{2+2d-n} + (d+1) \sqrt{(d-1)(n-2)} \right)^2}{2d^3(n-1)},
\end{equation}
which, as $d \to \infty$, evaluates to
\begin{equation}
F( \ket{y(k=n/2)}) = \frac{n-2}{2(n-1)}.
\end{equation}

This leaves only the $k=n$ case, which has two notable limits; $n=d$ and $n \ll d$.  For $n=d$ we find that
\begin{equation}
M = \frac{2}{d^3(d-1)} \left( \begin{array}{c c} 1 & 0 \\ 0 & 0 \end{array} \right)^{\oplus d(d-1)/2}.
\end{equation}
This makes determining the fidelity rather straightforward,
\begin{equation}
F( \ket{y(k=n=d)}) = \frac{d-1}{2d},
\end{equation}
which is equal to $1/2$ as $d \to \infty$.  In the $n \ll d$ case, we actually have that in the large $d$ limit,
\begin{equation}
\rho_1 \approx \frac 1d \sum_{i,j} \ketbra ij,
\end{equation}
which is pure, and therefore $\rho_2 = \rho_1 \otimes \rho_1$ and $F( \ket{y(k=n \ll d)}) = 1$.

The results of these calculations are somewhat surprising given our intuitions regarding monogamy constraints on the symmetric sharing of entanglement.  We had expected a heuristic connection between entanglement in the state and violation of the mean field approximation.  This notion was only partially correct though, as the mean field approximation is a stronger assumption than the separability of $\rho_2$.  Recall that a mixed state is separable if
\begin{equation}
\rho = \sum_i^r p_i \rho_i^{(1)} \otimes \rho_i^{(2)},
\end{equation}
for some decomposition, in which $r$ is unbounded.  The mean field approximation, however, demands $r=1$, which is therefore only true for a subset of separable states.  So it is then unsurprising that we were able to find PSS states for which $F(\ket \psi)$ was not close to 1, as the entanglement decaying with $n$ due to the symmetry implies that the state merely approaches a separable one, not one for which the mean field approximation is a necessarily good one.

The example of rectangular Young diagrams raises an important intuition regarding the validity of the mean field approximation.  The results of this section can be summarized as larger $k$ leading to better agreement with the mean field approximation.  Physically, small $k$ corresponds to more compact grouping of the particles.  Therefore we are led to believe that the more spread out the particles are, the better the mean field approximation gets.  This notion is given further context in the next section, where we conclude that the \emph{only} way to get good agreement with the mean field approximation is to have isolated particles.  This intuition does give hope to the use of the Gross-Pitaevskii Equation in quantum search.  The initial state for that algorithm is the uniform superposition,
\begin{equation}
\ket{\psi_0} = \left( \sum_{i=1}^d \ket i \right)^{\otimes n},
\end{equation}
which we know is approximately the $\ket{y(k=n)}$ state in the $n \ll d$ limit, and approaches perfect agreement with the mean field approximation.  Left to evolve, we would expect that the particles would stay mostly spread because that is both entropically and energetically favored.

\section{Bounded Fidelity Analysis}

The previous section introduced the intuition that isolated particles make for better agreement with the mean field approximation.  In this section we will confirm that notion by proving that good fidelity is \emph{impossible} without isolated particles.  The following theorem, whose proof can be found in the Appendix, will be instrumental in that endeavor,
\begin{thm} \label{fidbound}
In the limits that $1 \ll d$ and $n \sqrt n \ll d$, if a PSS state, $\ket{\psi}$, has $A \leq \mathcal O(d^{-2})$, then $F(\ket{\psi}) \leq 1/2$.
\end{thm}
This theorem allows us to identify any PSS state with $A \leq \mathcal O(d^{-2})$ as one for which the mean field approximation is not valid.  In finding sets of PSS states with such $A$, it will be important to establish notation which allows us to describe an arbitrary Young diagram, $y$, and its corresponding state vector, $\ket y$.  First, as before, label the number of rows as $k^{(y)}$, but now denote the number of distinct row lengths as $p^{(y)}$.  Denote the length of the $q^{\text {th}}$ distinct row from the bottom as $M_q^{(y)}$.  Denote the number of rows of length $M_q^{(y)}$ as $l_q^{(y)}$.  These labels are constrained by $p^{(y)}<k^{(y)}<d$ and $\sum_{q=1}^{p^{(y)}} l_q^{(y)}M_q^{(y)} =n$.  Finally, this notation allows us to determine the normalization coefficient, $\mathcal A_y$, for a single Young diagram basis element,
\begin{equation}
\mathcal A_y = \frac{d! \, n!}{(d-k^{(y)})! \Pi_y},
\end{equation}
where
\begin{equation}
\Pi_y = \prod_{q=1}^{p^{(y)}} l_q! \left[M_q^{(y)}! \right]^{l_q^{(y)}}.
\end{equation}

To confirm the intuition of the previous section, that isolated particles are required for good fidelity, let us start by considering the fidelity for \emph{single} basis element states.  In particular, let us examine Young diagrams which contain \emph{no} isolated particles, and denote the set of such Young diagrams as $\mathcal Y_>$,
\begin{equation}
\mathcal Y_> = \left \{ y \in \mathcal Y(n,d) \middle | M_1^{(y)} \geq 2 \right \},
\end{equation}
for example,
\ytableausetup{centertableaux,boxsize=6mm}
\[y \in \mathcal Y_> = \quad \quad
\begin{ytableau}
\; & & \none[\dots] & & & & \none[\dots] & & \\
\; & \none & \none & \none & \\
\none[\vdots] & \none & \none & \none & \none[\vdots] & \none & \none[\iddots] \\
\; & \none & \none & \none & \\
\tikznode{a4}{~} & & \none[\dots] & & \tikznode{a5}{~}
\end{ytableau}.
\]
\tikz[overlay,remember picture]{%
\draw[decorate,decoration={brace},thick] ([yshift=-3.5mm,xshift=3mm]a5.south east) -- 
([yshift=-3.5mm,xshift=-3mm]a4.south west) node[midway,below]{$\geq 2$};
}
\vspace{5mm}

We can then confirm that $A$ for any $\ket y$ such that $y \in \mathcal Y_>$ obeys $A \leq \mathcal O(d^{-2})$, and therefore, by Theorem \ref{fidbound}, $F(\ket y) \leq 1/2$.  To see this, start by performing the partial trace on $\ket y$ to find $\rho_1$, which amounts to finding $\mathcal N_{i,j}^{(y)}$.  Obviously, if $M_1^{(y)}=1$, there will be a contribution to $ \mathcal N_{i,j}^{(y)}$ which is proportional to $l_1^{(y)}$.  But for $M_1^{(y)} \geq 2$, the only way to contribute to $\mathcal N_{i,j^{(y)}}$ is if $i$ and $j$ are in row blocks $q$ and $q+1$, and $M_q^{(y)}+1=M_{q+1}^{(y)}$.  To add some intuition to that statement, we can only add to $ \mathcal N_{i,j}^{(y)}$ if, after removing a single block from $y$, there are at least two places (one for $i$ and one for $j$) to put that block back to return to $y$.  All that remains is to sum over the possible arrangements and selections of the remaining indices which construct an element in $y$,
\begin{align}
\mathcal N_{i,j}^{(y)} =& \frac{(d-2)!(n-1)!}{(d-k^{(y)})! \Pi_y} \biggr(\delta(M_1^{(y)}-1) l_1^{(y)}(d-k^{(y)}) \\ \notag &+ 2\sum_{q=2}^{p^{(y)}} \Delta^{(y)}(q,1) l_q^{(y)}l_{q-1}^{(y)}M_q^{(y)} \biggr),
\end{align}
where $\Delta^{(y)}(q,r) = \delta \left( M_q^{(y)}-M_{q-1}^{(y)}-r \right)$.  From here we can determine $A$, and find the following bounds,
\begin{align}
A =& \frac{2}{d^2n} \sum_{q=2}^{p^{(y)}} \Delta^{(y)}(q,1) l_q^{(y)} l_{q-1}^{(y)} M_q^{(y)} \\
\leq& \frac{2}{d^2n} \sum_{q=2}^{p^{(y)}} l_q^{(y)} l_{q-1}^{(y)} M_q^{(y)} \\
<& \frac{2k^{(y)}}{d^2n} \sum_{q=2}^{p^{(y)}} l_q^{(y)} M_q^{(y)} \\
<& \frac{2k^{(y)}}{d^2}.
\end{align}
So indeed, $A \leq \mathcal O(d^{-2})$ so long as $k^{(y)} = \mathcal O(1)$, and therefore $F(\ket{y}) \leq 1/2$ for $y \in \mathcal Y_>$.

Now let us broaden the picture by considering superpositions of basis elements with no isolated particles,
\begin{equation}
\ket{\psi_>} = \sum_{y \in \mathcal Y_>} a_{y} \ket{y}.
\end{equation}
From this state, tracing down to $\rho_1$ gives two components of $A$,
\begin{equation}
A= \sum_y A_>^{(y)} + \sum_{y \neq z} A_\times^{(y,z)},
\end{equation}
where, before defining them formally, the components can be described as $A_>^{(y)}$ being the contribution from the single $M_1^{(y)}>1$ basis elements, and $A_\times^{(y,z)}$ being the cross terms from different elements.  Now, in more detail, we can start with the familiar term,
\begin{align}
A_>^{(y)} = \frac{2 \left|a_y \right|^2}{d(d-1)n} \sum_{q=2}^{p^{(y)}} \Delta^{(y)}(q,1)l_q^{(y)}l_{q-1}^{(y)}M_q^{(y)} .
\end{align}
The new term in the calculation is the cross term, $A_\times^{(y,z)}$, but notably not all cross terms are going to appear in the partial trace.  Two Young diagrams, $y$ and $z$, will only contribute to $A_\times^{(y,z)}$ if $\ket{i k_2 \ldots k_n}$ is in $\ket y$ and $\ket{j k_2 \ldots k_n}$ is in $\ket z$ or vice versa.  This then implies that Young diagrams, $y$ and $z$, differ by only one block placement.  Possibly a clearer way to describe this is that removing a single block from $y$ and from $z$ will arrive at the same Young diagram, or, more precisely, they are connected to a common vertex with $n-1$ blocks in Young's lattice \cite{SUTER2002233}.  This leads us to define the `compatibility function', $G(y,z)$, which evaluates to 1 if $y$ and $z$ are connected to a common vertex with $n-1$ blocks in Young's lattice, and evaluates to 0 otherwise.  The size of $A_\times^{(y,z)}$ will then depend on the number permutations of the remaining $k_2 \ldots k_n$ which are consistent with $y$ and $z$.  To quantify this, let $m_1^{(y)}$ indicate the row cluster from which the block is taken in $y$ and moved to the row cluster $m_2^{(y)}$ in $y$ to create $z$.  We can analogously define $m_1^{(z)}$ and $m_2^{(z)}$.  We could then choose which diagram, $y$ or $z$, to index the sum over the $k_2 \ldots k_n$.  Rather than committing to one, we will do both simultaneously, relying on the following identity for compatible $y$ and $z$,
\begin{align}
\frac{l_{m_1}^{(y)}l_{m_2}^{(y)}M_{m_1}^{(y)}}{\Pi_y} &= \frac{l_{m_1}^{(z)}l_{m_2}^{(z)}M_{m_1}^{(z)}}{\Pi_z} \\  &= \sqrt{\frac{l_{m_1}^{(y)}l_{m_2}^{(y)}M_{m_1}^{(y)}}{\Pi_y}\frac{l_{m_1}^{(z)}l_{m_2}^{(z)}M_{m_1}^{(z)}}{\Pi_z}}.
\end{align}
From here we can finally determine
\begin{align}
A_\times^{(y,z)} = \frac{a_y^{\vphantom *}a_z^*}{d(d-1)n}G(y,z) \sqrt{l_{m_1}^{(y)}l_{m_2}^{(y)}M_{m_1}^{(y)}l_{m_1}^{(z)}l_{m_2}^{(z)}M_{m_1}^{(z)}}.
\end{align}
We can then bound the sums over these terms, starting with that over $A_>^{(y)}$
\begin{align}
\sum_y A_>^{(y)} &= \frac{2}{d(d-1)n} \sum_y \left | a_y \right |^2 \sum_{q=2}^{p^{(y)}} \Delta^{(y)}(q,1)l_q^{(y)}l_{q-1}^{(y)}M_q^{(y)} \\
&< \frac{2}{d(d-1)} \sum_y \left | a_y \right|^2 \sum_{q=1}^{p^{(y)}} l_q^{(y)} \\
&\leq \frac{n}{d(d-1)} \sum_y \left|a_y \right|^2 \\
&= \frac{n}{d(d-1)},
\end{align}
which is clearly still $\leq \mathcal O(d^{-2})$ so long as $n \ll d$.  We can then turn our attention to bounding the sum over $A_\times^{(y,z)}$,
\begin{align}
\sum_{y \neq z} A_\times^{(y,z)} &= \frac{1}{d(d-1)n} \sum_{y \neq z} a_y^{\vphantom *}a_z^*G(y,z) \\ \notag & \hspace{10mm} \times \sqrt{l_{m_1}^{(y)}l_{m_2}^{(y)}M_{m_1}^{(y)}l_{m_1}^{(z)}l_{m_2}^{(z)}M_{m_1}^{(z)}} \\
&< \frac{n}{2d(d-1)} \sum_{y \neq z} a_y^{\vphantom *}a_z^*G(y,z) \\
&= \frac{n}{4d(d-1)} \sum_{y \neq z} \left(a_y^{\vphantom *} a_z^* + a_z^{\vphantom *}a_y^* \right) G(y,z) \\
&\leq \frac{n}{2d(d-1)} \sum_{y \neq z} \left|a_y \right| \left | a_z \right|G(y,z) \\
&\leq \frac{n}{d(d-1)} \sum_y \left|a_y \right| \sum_{z \leq y} \left|a_z \right| G(y,z),
\end{align}
where $z \leq y$ if $ \left|a_z \right| \leq \left|a_y \right|$.  Continuing on,
\begin{align}
\sum_{y \neq z} A_\times^{(y,z)} &\leq \frac{n}{d(d-1)}\sum_y \left|a_y\right|^2 \sum_{z \leq y} G(y,z) \\
&< \frac{n}{d(d-1)} \sqrt{\frac n2} \sum_y \left| a_y \right|^2 \\
&= \frac{n}{d(d-1)} \sqrt{\frac n2},
\end{align}
which is likewise $\leq \mathcal O(d^{-2})$ so long as $n \sqrt n \ll d$.  These two together imply that $A \leq \mathcal O(d^{-2})$, and therefore, by Theorem \ref{fidbound}, $F\left( \ket{\psi_>} \right) \leq 1/2$.

\section{Discussion}

The cumulative conclusion of the work of this paper is that isolate particles are \emph{required} for good agreement with the mean field approximation.  Exactly how fidelity increases with the number of isolated particles, however, remains unknown.  Ideally we would be able to bound or approximate the fidelity as a function of the ratio of isolated particles to non-isolated.  To do so, though, new techniques will need to be developed to evaluate the trace of $\sqrt M$ for PSS states with $A> \mathcal O(d^{-2})$.

We have found that the mean field approximation is not in general appropriate for identical particles on the complete graph.  It then follows that the Gross-Pitaevskii is not in general a good approximation to the Schrodinger Equation for such systems, potentially unless the particles of the system remain relatively spread throughout the dynamics.  This analysis provides interesting context to the use of non-linear dynamics for quantum algorithms, but is merely an initial characterization.  Search algorithms, for example, require a marked site or set of marked sites, which breaks the site symmetry.  This analysis would then need to be repeated for states with bipartite site symmetry, whose basis elements would consist of pairs of Young diagrams.

This work was supported, in part, by NSF grant PHY-1620846.

\appendix

\section{Appendix:  Proof of Theorem 1}

\begin{proof}
Begin by noting that the normalization of $\rho_2$ implies that $C_1 \leq \mathcal O(d^{-1})$ and $C_2 \leq \mathcal O(d^{-2})$ for $i \neq j$.  This then constrains each of $\left|B_1\right|$-$\left|B_5 \right|$ by the positivity of $\rho_2$.  The simplest constraint comes from enforcing that the minors with a single $B_i$ and its associated diagonal elements is positive.  This gives 
\begin{align}
\left|B_1 \right| &\leq \mathcal O \left(d^{-2} \right) \\
\left|B_3 \right| &\leq \mathcal O \left(d^{-2} \right) \\
\left|B_2 \right| &\leq \mathcal O \left(d^{-\frac 32} \right) \\
\left|B_5 \right| &\leq \mathcal O \left(d^{-\frac 32} \right) \\ 
\left|B_4 \right| &\leq \mathcal O \left(d^{-1} \right).
\end{align} 
We know that, in the large $d$ limit, $A=dB_3 + 2 \Re \left (B_5 \right)$.  This tightens the constraint on $B_3$ to $\left|B_3 \right| \leq \mathcal O(d^{-5/2})$.  The final required constraint is a tighter bound on $B_1$, which will be achieved by considering the larger minor of $\rho_2$ defined by,
\begin{equation}
\ytableausetup{boxsize=1.5mm}
\rho_{\ydiagram{1,1}} = \sum_{i \neq j} \sum_{k \neq l} \left \{ \rho_2 \right \}_{ij,kl} \ketbra{ij}{kl}.
\end{equation}
We can then show that the following are eigenvectors of $\rho_{\ydiagram{1,1}}$,
\begin{align}
\ket{\lambda_1} =& \sum_{i \neq j} \ket{ij} \\
\ket{\lambda_2} =& \ket{12} + \ket{21} + \frac{d-4}{d-2}\sum_{i>2} \ket{2i} \\ \notag &+ \ket{i2} - \frac 2{d-2} \sum_{ i \neq j  >2} \ket{ij}.
\end{align}
Starting with $\ket{\lambda_1}$ we have,
\begin{equation}
\rho_{\ydiagram{1,1}} \ket{\lambda_1} = \sum_{i \neq j} \sum_{k \neq l} \left \{ \rho_{\ydiagram{1,1}} \right \}_{ij,kl} \ket{ij}.
\end{equation}
And therefore
\begin{align}
\bra{ij} \rho_{\ydiagram{1,1}} \ket{\lambda_1} &= \sum_{k \neq l} \left \{ \rho_{\ydiagram{1,1}} \right \}_{ij,kl} \\
&= 2 C_2 + 4(d-2) B_3 + (d-2)(d-3) B_1.
\end{align}
This implies that the associated eigenvalue is
\begin{equation}
\lambda_1 = 2 C_2 + 4(d-2) B_3 + (d-2)(d-3) B_1.
\end{equation}
Moving to $\ket{\lambda_2}$ we have,
\begin{align}
\rho_{\ydiagram{1,1}} \ket{\lambda_2} =& \sum_{i \neq j}\biggr(2 \left \{ \rho_{\ydiagram{1,1}} \right \}_{ij,12} +2 \frac{d-4}{d-2} \sum_{k>2} \left \{ \rho_{\ydiagram{1,1}} \right \}_{ij,k2} \\ \notag &-\frac{2}{d-2} \sum_{ k \neq l >2} \left \{ \rho_{\ydiagram{1,1}} \right \}_{ij,kl}\biggr) \ket{ij}.
\end{align}
Element by element we can confirm that
\begin{align}
\bra{12} \rho_{\ydiagram{1,1}} \ket{\lambda_2} &= 2 C_2 + 2(d-4) B_3 -2(d-3)B_1 \\
&= \bra{21} \rho_{\ydiagram{1,1}} \ket{\lambda_2},
\end{align}
and for $i>2$,
\begin{align}
\bra{i2} \rho_{\ydiagram{1,1}} \ket{\lambda_2} =& 2 B_3 + 2 \frac{d-4}{d-2} \left( C_2 + (d-3) B_3 \right) \\ \notag &- \frac2{d-2} \left( 2(d-3)B_3 + (d-3)(d-4) B_1 \right) \\
=& \bra{2i} \rho_{\ydiagram{1,1}} \ket{\lambda_2} \\
\bra{i1} \rho_{\ydiagram{1,1}} \ket{\lambda_2} =& 2 B_3 + 2 \frac{d-4}{d-2} \left(B_3 + (d-3) B_1 \right) \\ \notag &- \frac2{d-2} \left( 2(d-3)B_3 + (d-3)(d-4) B_1 \right) \\
=& 0 \\
=& \bra{2i} \rho_{\ydiagram{1,1}} \ket{\lambda_2},
\end{align}
and for $ i \neq j >2$,
\begin{align}
\bra{ij} \rho_{\ydiagram{1,1}} \ket{\lambda_2} = -\frac{2}{d-2} \left(2 C_2 + 2(d-4) B_3 -2(d-3)B_1 \right),
\end{align}
which implies that the associated eigenvalue is
\begin{equation}
\lambda_2 = 2 C_2 + 2(d-4) B_3 -2(d-3)B_1.
\end{equation}
To enforce that $\rho_{\ydiagram{1,1}} \geq 0$, we must have that $\lambda_1 \geq 0$ and $\lambda_2 \geq 0$.  As $d \to \infty$ this evaluates to the following constraints on $B_1$,
\begin{align}
B_3 &\geq 0& &\rightarrow& -\frac {2C_2}{d^2} - \frac{4B_3}{d} \leq &B_1 \leq \frac {C_2}d + B_3 \\
B_3 &\leq 0& &\rightarrow& -\frac{C_2}{d^2} \leq &B_1 \leq \frac{C_2}d,
\end{align}
which can be combined to $ \left | B_1 \right | \leq \max \left \{ \mathcal O ( d^{-3}), \mathcal O(B_3) \right \}$, which, in this case, gives $\left | B_1 \right | \leq \mathcal O(d^{-5/2})$.

With the magnitudes of the off-diagonal elements of $\rho_2$ so constrained, we can then turn to performing the matrix multiplication to find $M$.  Before doing so, however, we can examine the symmetries of $M$.  Note that by the singular value decomposition, $\sqrt{\rho_1}$ has the same symmetry as $\rho_1$.  We can then confirm that $\sqrt{\rho_1 \otimes \rho_1}$ has $V$ symmetry,
\begin{align}
\left \{ V \sqrt{\rho_1 \otimes \rho_1} V^{\dagger} \right \}_{ij,kl} =&  \left \{ \sqrt{\rho_1} \right \}_{V(i),V(k)} \left \{ \sqrt{\rho_1} \right \}_{V(j),V(l)} \\
=& \left \{ \sqrt{\rho_1} \right \}_{i,k} \left \{ \sqrt{\rho_1} \right \}_{j,l} \\
=& \left \{ \sqrt{\rho_1 \otimes \rho_1} \right \}_{ij,kl}.
\end{align}
The same is not true for the full $U$ symmetry, however, which we can see if we consider $U$ as the swap operator and the following family of entries,
\begin{eqnarray}
\left \{ U \sqrt{\rho_1 \otimes \rho_1} \right \}_{ij,il} &=&  \left \{ \sqrt{\rho_1} \right \}_{j,i} \left \{ \sqrt{\rho_1} \right \}_{i,l} \\
& \neq & \left \{ \sqrt{\rho_1} \right \}_{i,i} \left \{ \sqrt{\rho_1} \right \}_{j,l}.
\end{eqnarray}
It is true, however, that
\begin{eqnarray}
\left \{ U \sqrt{\rho_1 \otimes \rho_1} U^{\dagger}\right \}_{ij,kl} &=& \left \{ \sqrt{\rho_1} \right \}_{j,l} \left \{ \sqrt{\rho_1} \right \}_{i,k} \\
&=& \left \{ \sqrt{\rho_1 \otimes \rho_1}\right \}_{ij,kl}.
\end{eqnarray}
This finally allows us to show that
\begin{align}
V M V^{\dagger} =& V \sqrt{\rho_1\otimes \rho_1} V^{\dagger} V \rho_2 V^{\dagger} V \sqrt{\rho_1 \otimes \rho_1} V^{\dagger} \\
=&  \sqrt{\rho_1\otimes \rho_1} \rho_2 \sqrt{\rho_1 \otimes \rho_1},
\end{align}
and
\begin{align}
U M =& U \sqrt{\rho_1\otimes \rho_1} U^{\dagger} U \rho_2  \sqrt{\rho_1 \otimes \rho_1} \\
=&  \sqrt{\rho_1\otimes \rho_1} \rho_2 \sqrt{\rho_1 \otimes \rho_1},
\end{align}
with of course the same being true for a right application of $U$.  This confirms that both $M$ and its square root have the same symmetries as $\rho_2$, meaning the matrix multiplication to find $M$ reduces to finding a set of $B_1$-$B_2$, $C_1$, and $C_2$.  The calculation of the diagonal elements of $M$ then reduces under the assumption that $A \leq \mathcal O(d^{-2})$ to
\begin{align}
\left \{ M \right \}_{ii,ii} &= \frac{C_1}{d^2} + \mathcal O \left(d^{-\frac72} \right) \\
\left \{ M \right \}_{ij,ij} &= \frac{C_2}{d^2} + \mathcal O \left(d^{- \frac92} \right).
\end{align}
We can then constrain the elements of $\sqrt M$ by enforcing $\sqrt M^2 = M$.  After assigning a set of $B'$'s and $C'$'s to $\sqrt M$, those constraints resulting from the diagonal elements of $M$ evaluate to
\begin{align}\label{sMF}
\left \{ M \right \}_{ii,ii} =& C_1'^2+ d^2 B_2'^2 + dB_4'^2+ 2d \left |B_5' \right|^2  \\ \label{sMS}
\left \{ M \right \}_{ij,ij} =& 2C_2'^2 + d^2 B_1'^2 + d \left | B_2' \right |^2+ 4dB_3'^2 + \left | B_5' \right |^2
\end{align}
Then can be reformulated to the following inequalities,
\begin{align}
C_1' &\leq \frac{ \sqrt{C_1}}{d} \\
C_2' &\leq \frac{ \sqrt{C_2}}{\sqrt 2d}.
\end{align}
If we parametrize the diagonal elements of $\rho_2$ in the large $d$ limit as $C_1 = \cos^2 \theta/d$ and $C_2 = \sin^2 \theta/d^2$ for $i \neq j$, we can finally evaluate
\begin{align}
F(\ket{\psi}) &= \left(d C_1' + d^2 C_2' \right)^2 \\
&\leq \left( \frac{\cos \theta}{\sqrt d} + \frac{\sin \theta}2 \right)^2 \\
&\leq \frac 12.
\end{align}
\end{proof}

\bibliography{mybib}

\end{document}